\documentclass[12pt, a4paper]{article}
\usepackage{amsmath,amssymb,amsfonts}
\usepackage[colorlinks]{hyperref}
\usepackage{graphicx}

\begin{document}

\title{Production of GeV-scale heavy neutral leptons in 3-body decays.  Comparison with the PYTHIA approach}
\author{Volodymyr M. Gorkavenko, Yuliia R. Borysenkova
\\ and Mariia S. Tsarenkova\\
\it \small Faculty of Physics, Taras Shevchenko National University of Kyiv,\\
\it \small 64, Volodymyrs'ka str., Kyiv 01601, Ukraine}
\date{}

\maketitle
\setcounter{equation}{0}
\setcounter{page}{1}%

\begin{abstract}
Despite the undeniable success of the Standard Model of particle physics (SM), there are some phenomena that the SM can't explain.
These phenomena indicate that the SM has to be modified. One of the possible ways to extend the SM is to introduce heavy neutral leptons (HNLs). To search for HNLs in intensity frontier experiments, one has to consider HNL production both in 2-body and 3-body decays of some mesons. We verified the possibility of using the parton level PYTHIA  default matrix elements (without the form-factor formalism) to calculate HNL production in 3-body semileptonic decays of $B$ and $D$ mesons in the experimentally interesting mass range of the produced HNLs.   We conclude that this approach is quite suitable for the estimation of the sensitivity region for HNLs in the intensity frontier experiments, provided one uses suitable parton level PYTHIA default matrix elements. Our study was driven by the usage of such an approximation by the SHiP collaboration. 
We conclude that in this case the parton level PYTHIA default matrix elements could have been chosen more appropriately.

\end{abstract}

\section{Introduction}\label{intro}
The Standard Model of particle physics (SM) \cite{SM1,SM2,SM3} is a theory that describes with high precision the processes of electroweak and strong interactions with the participation of elementary particles. It is consistent up to a very high energy scale (perhaps up to the Planck scale) and it is verified in numerous accelerator experiments up to energy $\sim 15$ TeV. However, the SM fails to explain some phenomena such as massiveness of neutrinos (see e.g. \cite{Bilenky1, Salas}),  dark matter (for reviews see e.g. \cite{Peebles,Lukovic,Bertone}), dark energy \cite{Brax}, baryon asymmetry of the Universe \cite{Steigman}, etc. Therefore, the SM is an incomplete theory and it requires an extension. One has to suggest the existence of "hidden" sectors with particles of new physics.

It turns out that the mentioned SM problems can be theoretically solved by extending the SM by new particles that can be either heavy or light. Indeed, neutrino oscillations and the smallness of the active neutrino masses can be explained with the help of new particles with sub-eV mass as well as with the help of heavy particles of the GUT scale, see e.g. \cite{Strumia}. The same may be said about the baryon asymmetry of the Universe and dark matter problems: physics at very different scales can be responsible for them, see e.g.  \cite{kolb}.

So, two possible answers can be formulated to the question "why do we not observe particles of new physics in experiments?" The first answer is the following.   The new particles are very heavy and can't be produced in modern accelerators like the LHC. To detect them one has to build more powerful and more expensive accelerators (we need energy frontier experiments).
However, there is another possibility. 
The particles of new physics can be light (with a mass below or of the order of the electroweak scale) that feebly interact with the SM particles. The last case is very interesting for the experimental search for new physics right now, see e.g. \cite{Gorka}.  To search for rare interactions of feebly interacting hypothetical particles, the intensity frontier experiments are needed. These experiments aim to create high-intensity particle beams and use large detectors \cite{Beacham}. Several such
intensity frontier experiments have been proposed in recent years: DUNE \cite{LBNE}, NA62 \cite{Mermod,NA62,Drewes}, SHiP \cite{Alekhin,SHiP}, etc.

We do not know what are the properties of the new particles. They can be new scalars, pseudoscalars, vectors or fermions, see \cite{Alekhin,MatPhysCase} for a review. Each of these options has to be tested in experiments. From a theoretical point of view,
there are three possible choices of the new renormalized Lagrangian of the interaction of new particles  with the SM particles. These interactions are called portals. 
There are scalar  (e.g. \cite{Patt,Bezr,Boiar}), vector (e.g. \cite{Okun,Bob,Langacker}) and heavy neutral leptons renormalized portals. It means the new interaction can be observed at any energy scale, including that below the EW scale. There are also other portals of high-dimensional operators such as the portal of pseudoscalar particles (axion-like particles), see e.g. \cite{Peccei, Weinberg, Wilczek} and \cite{Kiwoon} for a review, or Chern-Simons like (parity odd) interaction of electroweak gauge bosons with a new vector field (e.g. \cite{Anastasopoulos,Antoniadis}). The lower the energy scale is, the less important these interactions will be.

In this paper, we consider extending the SM by neutrino singlets with right chirality, which extremely faintly
interact with the SM particles. Such right-handed neutrinos are called sterile
neutrinos or heavy neutral leptons (HNLs).

Interest in the HNL modification of the SM is conditioned by the model's ability to explain the smallness of active 
neutrino masses (due to large values of the sterile neutrino masses $M_I$ or small values of Yukawa elements $F_{\alpha I}$) and to describe the generation of the matter-antimatter
asymmetry of the Universe due to CP violation in the model \cite{Fukugita}. 
It was shown that Majorana masses of HNLs can be GeV scale \cite{Akhmedov, Shap1, Shap3} to explain baryon asymmetry of the Universe. 

In 2005 the Neutrino Minimal
Standard Model ($\nu MSM$) model was proposed \cite{Shap1, Shap2}. 
In this model, the SM is extended by three right-handed neutrinos (heavy neutral leptons)
with masses smaller than the electroweak scale. It was shown that 18 new parameters of the
model can be chosen in such a way as to simultaneously solve problems of neutrino oscillations, baryon asymmetry in the Universe, and dark matter. 
In this case, the $\nu MSM$ model requires the existence
of two right-handed neutrinos with  practically
the same masses ($\gtrsim  100$ MeV) and one right-handed neutrino
with a relatively small mass in the keV region, see \cite{nuMSMR} for a review. 
The lightest right-handed neutrino is a long-lived particle, a dark matter candidate. In 2014 the possible manifestation
of the lightest  right-handed neutrino with mass 7 keV was found in  X-ray spectra of the Andromeda
galaxy and the Perseus galaxy cluster \cite{Yakub, Bulbul}. 

That is why the HNL extension of the SM attracts a lot of attention and interest. 
This modification of the SM is especially interesting for GeV-scale HNLs that can be in principle detected in the intensity frontier experiments \cite{Beacham,Alekhin}. 

A model-independent phenomenological approach is used in the experimental search for HNLs, assuming the existence of only one HNL and considering that the other HNLs do not affect the analysis. For simplicity of analysis, restrictions are imposed on only two free parameters of the model: Majorana mass of the appropriate HNL ($m_N$) and the mixing angle of the interaction of this HNL with only one active neutrino of flavour $\alpha$ ($U_{\alpha}$). While it is usually assumed that the mixing angles of interaction with other active neutrinos are zero. 

It should be noted that the results of previous numerous experiments almost completely closed the region for HNLs with masses below the mass of kaon, see \cite{Alekhin, Bondar} for details. Therefore,  HNLs with mass $m_N\gtrsim 0.5$ GeV are of interest to us.
The phenomenology of GeV-scale HNLs was considered e.g. in \cite{Bondarenko:2018ptm}. A recent computation of the sensitivity region for HNL search in the SHiP experiment was performed in \cite{Ahdida}.

\begin{figure}[t]% figure* for wide figure, [h] [!] to change the placement
\includegraphics[width=\textwidth]{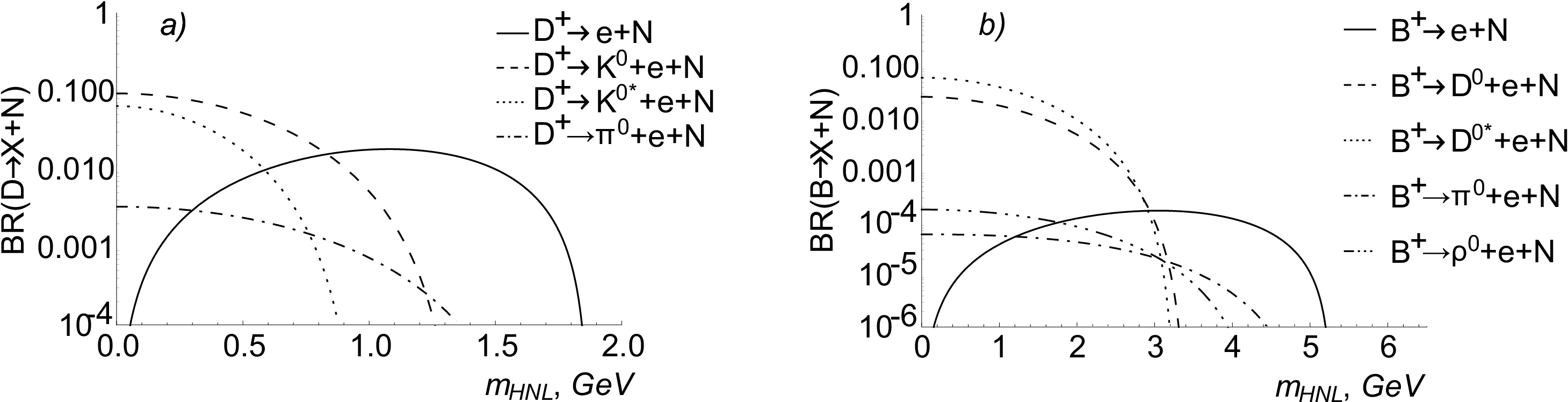}
\caption{Dominant branching ratios of HNL production from charged $D$ (figure \textit{a})  and $B$ (figure \textit{b}) mesons, see \cite{Bondarenko:2018ptm} for details. Here $U_e = 1$,
$U_\mu = U_\tau = 0$.
}\label{fig:BrD}
\end{figure}

If we consider the most important production channels of HNLs in the fixed target intensity frontier experiments (such as NA62, SHiP or DUNE),  we see that they are the semileptonic decays of $D$, $D_s$ mesons,  $B$, $B_s$, $B_c$ mesons and decays of $\tau$ leptons, see \cite{Bondarenko:2018ptm}. It should be noted that an essential contribution to HNL production is given by 3-body decays of the mentioned particles, see figure \ref{fig:BrD}, 
where we present branchings for decays of $D$ and $B$ mesons as an example. Branchings for decays of other mesons can be found in \cite{Bondarenko:2018ptm}.

Taking into account that the number of the produced $D$ mesons is sufficiently greater than the number of the produced $B$ mesons in proton-target collisions (e.g. $\sim 10^{17}$ and $\sim 10^{13}$ correspondingly for 5 years of SHiP experiment operation \cite{Bondarenko:2018ptm}), it is obvious that HNLs production from $B$ mesons decay can be neglected for HNLs with masses $m_N \lesssim 2$ GeV.

So, 3-body decays of mesons with $\tau$ leptons in the final state are either forbidden (for the decays of $D$ mesons) or ineffective (for the decays of $B$ mesons). We will denote the final states of charged leptons as $\ell=e,\mu$.

In the SHiP collaboration paper \cite{Ahdida} computation of the 3-body semileptonic decay contributions ($h \rightarrow h^\prime +\ell + N$) for the formation of the sensitivity region for HNLs was based on PYTHIA 8.
PYTHIA is a general purpose collision event generator \cite{Pythia},  that can be modified to embody new simulations with an arbitrarily good approximation in principle.
However, the 3-body decay contributions to the sensitivity region built in \cite{Ahdida} were computed with use of the parton level PYTHIA default matrix elements without additional tuning. This can be clearly seen by looking at their program code\footnote{The calculations of the SHiP collaboration were carried out with the use of the FairShip software framework. The program code is in open access, see \mbox{https://github.com/ShipSoft/FairShip }}. 
It motivated us to  verify the accuracy of this approach.

The fact is that for the description of the semileptonic decays of $D$ mesons PYTHIA uses a default matrix element %\vspace{-0.7em}%\newpage
\begin{equation}\label{PythD}
    |M_{fi}|^2=(p_h p_\ell)(p_\nu p_{h^\prime})
\end{equation}
and for description of the semileptonic decays of $B$ mesons PYTHIA uses a default matrix element
\begin{equation}\label{PythB}
    |M_{fi}|^2=(p_h p_\nu)(p_\ell p_{h^\prime}).
\end{equation}
Explicit form of these matrix elements is given in the PYTHIA 6.4 manual \cite{Pythia64}. The current version, PYTHIA 8, uses the same matrix elements without change. Our further computations are valid for PYTHIA 6 as well as for PYTHIA 8. 

However, for accurate consideration we need to use at least form-factors of meson transitions.
In this paper, we analyze the relevancy of the usage of the parton level default PYTHIA matrix elements in the SHiP collaboration paper  \cite{Ahdida}  for the
computation of the contributions of the 3-body decays of $B$ and $D$ mesons to the sensitivity region.

As it was pointed out in \cite{Alekhin}
HNLs can be produced in $\tau$ lepton decays and these decays are important in case of dominant mixing with the $\tau$ flavour.
The main 3-body decay channels of $\tau$ leptons are decays into elementary particles $\tau\to N \ell_\alpha \bar{\nu}_\alpha$ and $\tau\to \nu_\tau \ell_\alpha N$, where $\alpha=e,\mu$. 
In contrast to meson decays, which must be described using the form-factor formalism, these decays can be directly  described by the PYTHIA formalism. 

The paper is organized in the following  way.
In section \ref{sec:neutr} we consider a general formalism of the neutrino modification to the Standard Model.
In section \ref{sec:strategy} we formulate a
general strategy  to get the domain of parameters that allows  hidden particles
to be detected in the intensity frontier experiments. 
In section \ref{sec:pdf_def} we get the general definition of a probability density function (PDF) for the production of particles  with a certain value of energy  in 3-body decays.
In section \ref{decay:pseudoscalar} and section \ref{decay:vector} we present the exact matrix elements for the 3-body decays of mesons.
In section \ref{sec:pdf_hnl} we compare PDFs for HNL production in 3-body decays of mesons (in the own reference frame of the mesons) computed using the parton level PYTHIA default matrix elements and the exact ones.
In section \ref{sec:dist_funcs} we derive the energy and polar angle distribution functions of the produced HNLs in the laboratory reference frame.
In section \ref{sec:egeom} we find the probability of a produced
HNL to fall on the detector.
In section \ref{sec:pdecay} we estimate the probability of
a produced HNL to decay inside  the vacuum tank before
the detector.
Finally, the results are summarized in section \ref{conclusions}.  
Useful kinematic relations  for 
HNLs in the different reference frames are
outlined in appendix \ref{appendix}.

\section{Neutrino modification of the SM. Heavy Neutral Leptons\label{sec:neutr}}

Renormalized interaction of the right-handed neutrinos with the SM particles (HNL portal)
is similar to the Yukawa interaction of left-handed quark doublets
with singlets of the right-handed quarks in the SM, namely:
\begin{equation}\label{sh0}
\mathcal L_{int}=-\left(F_{\alpha I}\bar L_\alpha \tilde H
N_I+\textrm{h.c.}\right),
\end{equation}
where  $\alpha=e,\mu,\tau$, the index $I$ 
is from 1 to the full number of the sterile neutrinos ($n$),  $L_{\alpha}$ -- the
doublet of the leptons of $\alpha$-generation, $N_I$ -- a right-handed sterile neutrino,
$F_{\alpha
I}$ -- a new matrix of dimensionless Yukawa couplings, ${\tilde H}={\rm i}\sigma_2H^*$.

Conditions of invariance of the Lagrangian \eqref{sh0} to the transformations of the gauge groups of the SM demand the corresponding charges of the sterile neutrinos to be zero. Therefore, sterile neutrinos are not charged relative to the gauge groups of the SM, which justifies their name.

After the electroweak symmetry breaking, Lagrangian \eqref{sh0} in the unitary gauge looks as follows: 
\begin{equation}\label{sh0Un}
\mathcal L_{int}=-\left(M^D_{\alpha I}\bar \nu_\alpha   
N_I+\textrm{h.c.}\right),  \quad M^D_{\alpha I} = \frac{v}{\sqrt{2}}F_{\alpha I},
\end{equation}
where $v\approx 246$ GeV is the vacuum expectation value of the Higgs field, $M^D_{\alpha I}$ --  Dirac mass terms.

Considering sterile neutrinos as neutral Majorana particles, we can write the full Lagrangian of the modified neutrino sector of the SM in the form
\begin{equation}
\label{sh0Neutri}
\mathcal L_{\nu,N}={\rm i} \bar{\nu_k} \not{\!\partial} \nu_k + {\rm i} \bar{N_I} \not{\!\partial} N_I-
  \left(M^D_{\alpha I}\bar{\nu}_{\alpha} N_I + \frac{M_{I}}{2} \bar{N_I}^c N_I + \textrm{h.c.}\right),
\end{equation}
where $M_I$ is the Majorana mass of $I$th sterile neutrino. 
Imposing the condition $M^D_{I\alpha}/M_I\ll 1$, one can perform the diagonalization of the neutrino mass matrix, see e.g. \cite{Strumia,Bilenky}, and get the mass matrix of the active neutrinos
\begin{equation}\label{Mact1}
(M_{\nu}^{active})_{\alpha\beta}=-\sum_{I=1}^n\frac{M^D_{
I\alpha}M^D_{I \beta}}{M_I}.
\end{equation}
The mass matrix for the sterile neutrinos 
will remain almost unchanged.  This mechanism is known as the seesaw mechanism\footnote[1]{This mechanism is  called Type-I seesaw mechanism because there are other ways to explain small neutrino masses, see e.g. \cite{Strumia, Ma}. In the Type-II seesaw mechanism
an extra SU(2) triplet scalar is introduced \cite{Lazarides,Mohapatra1,Schechter,Ma1}, in the type-III seesaw mechanism an extra
fermion in the adjoint of SU(2) is added to the model \cite{Foot}.}, see e.g. \cite{Mohapatra, Yanagida}.

As a result of the neutrino states mixture, the active neutrino states become superposition of the mass states of the active and sterile neutrinos 
\begin{equation}\label{nuaktive}
    \nu_L=\left(1-\frac{1}{2}U^+U\right)V_1\nu_{mL}+U^+V_2 N^c,
\end{equation}
where $U_{I\alpha}=M^D_{I\alpha}/M_I$ ($U=M^{-1}M^D$) is a so-called mixing angle ($U_{I\alpha}\ll 1$), $V_1$ -- a unitary matrix for diagonalization of the active neutrino mass matrix, $V_2$ -- a unitary matrix for diagonalization of the sterile neutrino mass matrix (it can be taken as a unit matrix in our case). It means that sterile neutrinos interact with the SM particles similarly to active neutrinos:
\begin{equation}\label{WZNinteraction}
      \mathcal{L}_{int}\! =\! -\frac{g}{2\sqrt 2} W^+_\mu\sum_{I,\alpha} \overline{N^c}_I  U_{I\alpha} \gamma^\mu (1-\gamma_5) \ell^-_\alpha\! -\! \frac{g}{4 \cos\theta_W}Z_\mu \sum_{I,\alpha}\overline{N^c}_I  U_{I\alpha} \gamma^\mu (1-\gamma_5) \nu_\alpha + \mbox{h.c.}
\end{equation}
It should be noted that the extension of the SM by HNLs gives an additional source of CP-symmetry violation in the theory, see e.g. \cite{Gorka10}.

\section{General strategy}\label{sec:strategy}

\begin{figure}[t]% figure* for wide figure, [h] [!] to change the placement
\centering
\includegraphics[width=0.7\textwidth]{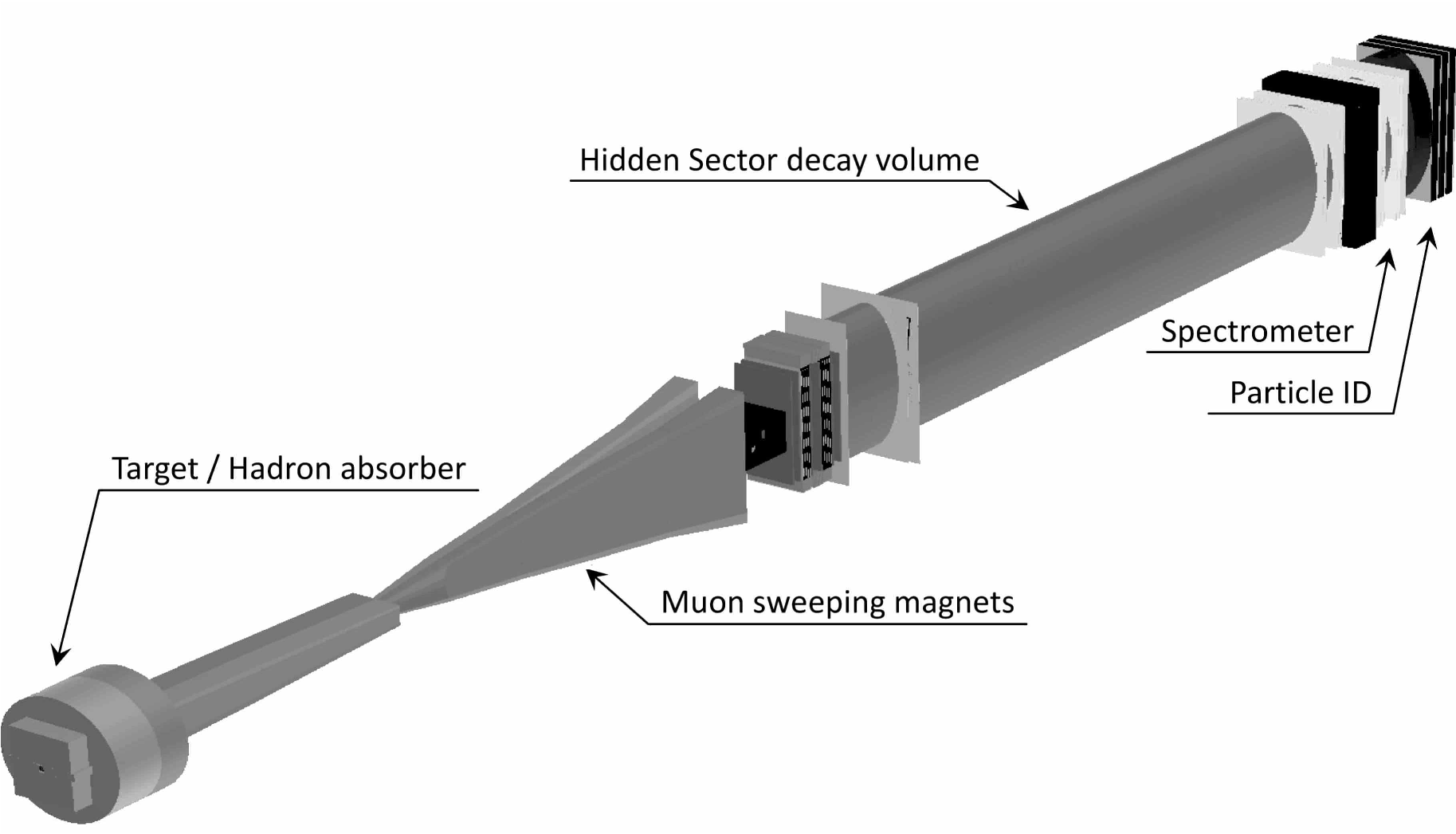}
\caption{General scheme of the SHiP facility.}\label{Ship}
\end{figure}

At first, let us remind the general principles of the intensity frontier
experiments operation on the example of an intended experiment SHiP \cite{Ahdida1}, see figure \ref{Ship}.
A beam line from the CERN SPS accelerator will transmit 400 GeV protons at the SHiP. 
A proton beam will strike in a Molybdenum and Tungsten fixed target at a
center-of-mass energy $E_{CM} \approx 27$ GeV. 
A great number of light SM particles and hadrons will be produced under such collisions. 
Hidden  particles are expected to be predominantly produced in the decays of the produced hadrons. 

The main concept of the SHiP functioning is the following. Almost all the produced SM particles should be either trapped by an absorber or deflected in a magnetic field (muons). Remaining events with the SM particles can be rejected using specially developed cuts.
If hidden particles decay into SM particles inside the decay volume, the latter will be detected. It will mean the existence of hidden particles.

We can estimate the number of hidden particles that can be detected as
\begin{equation}\label{Ns0}
N_{det}=N_{BSM}\cdot \epsilon_{tot} \cdot P_{decay}.
\end{equation}
This relation  allows us to find a range  of parameters  $(m_N,\theta_{\alpha})$,
when HNLs can be detected.  It is a region where $N_{det}\geqslant N_{det}^{thr}$. 
Value  $N_{det}^{thr}$ is the threshold number of the detected particles when we can 
affirm the discovery of HNLs.  It depends on the characteristics of the experiment facilities and the background level.

Let us explain the meaning of the constituent factors in \eqref{Ns0}.
Factor $N_{BSM}$ is the number of hidden particles produced  during all the time of the SHiP experiment
operation. 
Factor $\epsilon_{tot}$ is 
the product of the following factors $\epsilon_{geom}\cdot  Br_{vis} \cdot \epsilon_{det}$, where $\epsilon_{geom}$ is the probability of a produced HNL
to move towards the detector, $\epsilon_{det}$ is the probability of the detector to register visible particles,  $Br_{vis}$ is the branching of HNL decay into channels, visible for the detectors. Factor $P_{decay}$ is the probability of a
produced HNL to decay in the volume of the vacuum tank
before the detectors.

Therefore, to validate the accuracy of the computation with use of the parton level PYTHIA default matrix elements  of the contribution of the produced in the 3-body decays HNLs to the sensitivity region  we have to consider only two  factors in \eqref{Ns0}, namely $\epsilon_{geom}$  and $P_{decay}$.

The probability for the produced HNLs to move towards the
detector ($\epsilon_{geom}$) can be easily found if we know the
distribution function of these  particles. The probability of the
 HNLs to decay inside the vacuum tank before the detector is
\begin{equation}\label{a31}
P_{decay}= e^{-\frac{L \Gamma}{\gamma \beta} } -
 e^{-\frac{(L+\Delta L) \Gamma}{\gamma \beta} },
\end{equation}
where $L$ is the distance from the target to the vacuum tank, $\Delta
L$ -- the length of the vacuum tank, $\gamma$ is the Lorentz factor,
$\beta$ -- the velocity of the particle and $\Gamma$ -- the decay width
of HNLs. Thus,
$P_{decay}$ depends on the energy distribution function of HNLs, HNL's
lifetime, geometry of the experiment, and the coupling constant.

Therefore, to compute both $\epsilon_{geom}$ and $P_{decay}$ we need the energy-angle distribution functions of the produced HNLs. One can easily get these functions in the rest frame of the initial meson. Using the energy-angle distribution of the initial mesons, we can get the distribution functions of HNLs in the laboratory reference frame. We  assume that in the initial meson's rest frame the production of HNLs is isotropic.

\section{Probability density function for particles produced in a 3-body decay}\label{sec:pdf_def}

Let us consider a 3-body decay $A\rightarrow B + C +N$, where $A$, $B$, $C$ are some particles (with masses $m_A$, $m_B$, $m_C$) and $N$ is a sterile neutrino with mass $m_N$.

If the decaying particle ($A$) is a scalar (or we average over its spin states), the 
differential decay width of the 3-body decay in the rest frame of the $A$ particle is defined as, see e.g. \cite{PDG}, 
\begin{equation}\label{dalits10}
d\Gamma=\frac{\left|M_{fi}\right|^2}{8m_A(2\pi)^3}\,d E_N\,d E_B.
\end{equation}
The full partial decay width for this channel is given as
\begin{equation}\label{dalits10int}
\Gamma(A\rightarrow BCN)=\hspace{-0.5em}\int\limits_{E_{N}^{min}}^{E_{N}^{max}}\hspace{-0.8em} d E_N\hspace{-0.8em}
\int\limits_{E_{B}^{min}(E_N)}^{E_{B}^{max}(E_N)}\hspace{-1.6em} d E_B\,
\frac{\left|M_{fi}\right|^2}{8m_A(2\pi)^3},
\end{equation}
where  
the boundaries of integration can be found from  condition
\begin{equation}\label{dalits13}
 E_{N}^{2\,\,max/min }=m_N^2+\vec p_{N} {}^{2\,\,max/min }
=m_N^2+\vec p_B{\,}^2+\vec p_C{\,}^2\pm 2|\vec
p_B|\,|\vec p_C|,
\end{equation}
that can be rewritten as the solution of equation
\begin{eqnarray}
 E^{2\,\,max/min}_{N }=m_N^2+E_B^2-m_B^2+(m_A-E_N-E_B)^2-\nonumber\\\label{dalits13a}
-m_C^2\pm 2\sqrt{(E_B^2-m_B^2)((m_A-E_N-E_B)^2-m_C^2)},
\end{eqnarray}
namely
\begin{equation}\label{maxmin}
 E^{max/min}_{B}(E_N)=   \frac{(m_A-E_N)(w^2+m_B^2-m_C^2)}{2w^2}\pm\sqrt{E_N^2-m_N^2}\,\frac{\sqrt{\lambda(w^2,m_B^2,m_C^2)}}{2w^2},
\end{equation}
where $w^2 = m_A^2 + m_N^2 - 2E_N m_A$ and 
$\lambda (x,y,z) = x^2 + y^2 + z^2 - 2xy - 2yz - 2zx
$ is the K\"all\'en function \cite{Kallen}.

\begin{figure}[t]% figure* for wide figure, [h] [!] to change the placement
\centering
\includegraphics[width=0.9\textwidth]{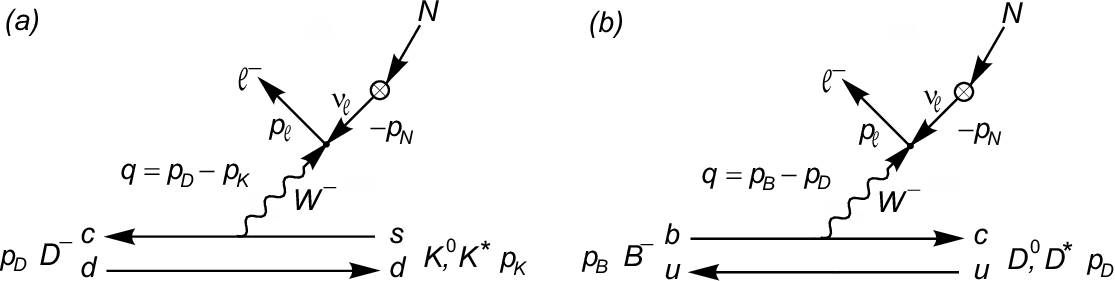}
\caption{Diagram of the semileptonic decays: (\textit{a}) -- decay of pseudoscalar meson $D^-$ into pseudoscalar meson $K^0$ or vector meson $K^*(892)$, (\textit{b}) -- decay of pseudoscalar meson $B^-$ into pseudoscalar meson $D^0$ or vector meson $D^*(2007)^0$.}\label{fig:diagps}
\end{figure}

As one can see, two functions $E_{B\,\,max/min}(E_N)$ in \eqref{maxmin} define the upper and lower boundaries
of the region with the allowed values of energy. These functions coincide at  
the minimal and maximum values of the energy $E_N$ that can be found from the condition for the changing sign term being zero. 
We get
\begin{equation}\label{ENminmax}
    E_{N}^{min}=m_N,\,\, E_{N}^{max}=\frac{m_A^2\! +\!m_N^2\!-\! (m_B\! +\! m_C)^2}{2 m_A}.
\end{equation}

Taking into account \eqref{dalits10int}, we get the probability density function (PDF) for the production of HNLs with
a certain value of energy ($E_N$) in the form
\begin{equation}\label{pdf_def}
    pdf(E_N)=  \frac{1}{\Gamma(A\rightarrow BCN)}\hspace{-1em}
\int\limits_{E_{B}^{min}(E_N)}^{E_{B}^{max}(E_N)}\hspace{-1.6em} d E_B\,
\frac{\left|M_{fi}\right|^2}{8m_A(2\pi)^3}.
\end{equation}

\section{HNL production in semileptonic decays of $B$ and $D$ mesons into pseudoscalar mesons}\label{decay:pseudoscalar}

Decay of  an electrically charged pseudoscalar meson $h$ into an electrically neutral pseudoscalar meson $h^\prime$, a charged lepton and an HNL	($h \rightarrow h^\prime +\ell + N$) is derived by weak interaction, see figure \ref{fig:diagps}. 
The amplitude of the reaction is
\begin{equation}\label{Mfi_Benu}
 M_{fi}=\theta_\alpha \frac{G_F}{\sqrt2}\,V_{ij}^*\,\bar \ell_\alpha \gamma^\nu(1-\gamma^5) N \langle h^\prime(p^\prime) |\bar Q_i\gamma_\nu(1-\gamma^5)Q_j |{h(p)}\rangle,
\end{equation}
where  averaging over the axial quark current  gives zero, but 
averaging over vector quark current can be presented as
\begin{equation}\label{qcuraveraged}
  \mathcal{W}_\nu = \langle h^\prime(p^\prime) |\bar Q_i\gamma_\nu Q_j |{h(p)}\rangle =
\!\left[\!(p\!+\!p^\prime)_\nu\! -\! \frac{m_h^2\! - \!m_{h^\prime}^2}{q^2} q_\nu\! \right]\!f^{hh^\prime}_+\!(q^2)
\! +\!\frac{m_h^2\! -\! m_{h^\prime}^2}{q^2} q_\nu f^{hh^\prime}_0\!(q^2), 
\end{equation}
where $f^{hh^\prime}_+(q^2)$ and $f^{hh^\prime}_0(q^2)$ are the form-factors for ${hh^\prime}$ transitions.

For $|M_{fi}|^2$ summarized over the helicities of the final particles, we have
\begin{equation}\label{mfiFinalN} 
\overline{|M_{fi}|^2} = 4 \theta_\alpha^2  G_F^2 |V_{ij}|^2  [\vphantom{(\mathcal{W})^2}
 2(m_{h}^2 - m_{h^\prime}^2) f^{hh^\prime}_0(q^2) \left(k \mathcal{W}\right)
 - 2\left(k \mathcal{W}\right)^2 - (qk)\mathcal{W}^2 + m_N^2\mathcal{W}^2\,],
\end{equation}
where $m_{h}$, $m_{h^\prime}$ are the masses of the corresponding mesons, $m_N$ -- the mass of the sterile neutrino.

For consideration of reaction $B^\pm \rightarrow D^0 + \ell^\pm +N$ we use a popular parametrization for form-factors of mesons, namely the Bourrely-Caprini-Lellouch (BCL) parametrization~\cite{Bourrely} that takes into account the analytic properties of the form-factors (see e.g.~\cite{Aoki,Na}),
\begin{equation}\label{eq:27}
   f(q^2) = \frac{1}{1-q^2/{M_{\rm pole}^2}} \sum_{n=0}^{N-1} a_n\biggl[\bigl(z(q^2)\bigr)^n - (-1)^{n-N} \frac nN \bigl(z(q^2)\bigr)^N\biggr],
\end{equation}
where
function $z(q^2)$ is defined via
\begin{equation}
  \label{eq:28}
  z(q^2) \equiv\frac{\sqrt{t_+ - q^2} - \sqrt{t_+ - t_0}}{\sqrt{t_+ - q^2} + \sqrt{t_+ - t_0}}
\end{equation}
with
\begin{equation}
  \label{eq:16}
  t_+  = \bigl(m_h + m_{h'}\bigr)^2.
\end{equation}
The choice of $t_0$ and the pole mass $M_{\textrm{pole}}$ varies from
group to group that performs the analysis. In this work we follow the FLAG collaboration~\cite{Aoki} and take
\begin{equation}
 t_0 = \bigl(m_h + m_{h'}\bigr) \bigl(\sqrt {m_h} - \sqrt{m_{h'}}\bigr)^2.
 \label{eq:t0FLAG}
\end{equation}
The coefficients $a_n^+$ and $a_n^0$ are then fitted to the experimental data or lattice results.  Their best fit parameter values are given in table  \ref{tab:Bscalarformfactors}.

\begin{table}[t]
\noindent\caption{Best fit parameters for the form-factors~\eqref{eq:27} of  $B\to D$ transitions~\cite{Aoki}.}\label{tab:Bscalarformfactors}\vskip3mm\tabcolsep4.5pt
\centering
\begin{tabular}{|c|c|c|c|c|}
\hline
$f$ & $M_{\textrm{pole}}$~(GeV) & $a_0$ & $a_1$ & $a_2$ \\ \hline
$f_+^{BD}$ & $\infty$ & $0.909$ & $-7.11$ & $66$ \\ \hline
$f_0^{BD}$ & $\infty$ & $0.794$ & $-2.45$ & 33 \\ \hline
\end{tabular}
\end{table}

For consideration of reaction $D^\pm \rightarrow K^0 +\ell^\pm + N$ we use the parametrization for form-factors of mesons given in \cite{ETM}:
\begin{equation}
 f(q^2) = \frac{f(0) - c(z(q^2)-z_0)\left(1+\frac{z(q^2)+z_0}{2}\right)}{1 - P q^2},
 \label{eq:fLubicz}
\end{equation}
where $z(q^2)$ is defined by \eqref{eq:28} and $z_0=z(0)$. The best fit parameter values are given in table  \ref{tab:Dscalarformfactors}.

\begin{table}[t]
\noindent\caption{Best fit parameters for the form-factors~\eqref{eq:fLubicz} of $D\to K$ transitions~\cite{ETM}.}
\label{tab:Dscalarformfactors}\vskip3mm\tabcolsep4.5pt
\centering
\begin{tabular}{|c|c|c|c|}
\hline
$f$ & $f(0)$ & $c$ & $P~(\textrm{GeV}^{-2})$  \\ \hline
$f_+^{DK}$ & $0.7647$ & $0.066$ & $0.224 $ \\ \hline
$f_0^{DK}$ & $0.7647$ & $2.084$ & $0$ \\ \hline
\end{tabular}
\end{table}

\section{HNL production in semileptonic decays of $B$ and $D$ mesons into vector mesons}\label{decay:vector}

Let us consider production of HNLs in semileptonic  decays of $B$ and $D$ mesons into vector mesons, namely 
$B^\pm \rightarrow D^*(2007)^0 +\ell^\pm + N$ and  $D^{\pm} \rightarrow K^*(892) + \ell^{\pm} + N$. 

Decay of an electrically charged pseudoscalar meson $h$ into an electrically neutral vector meson $h^\prime_V$, a charged lepton and an HNL	($h \rightarrow h^\prime_V +\ell + N$) is derived by weak interaction, see figure \ref{fig:diagps}. 
The amplitude of the reaction is similar to \eqref{Mfi_Benu},
where  there is contribution from both the vector
and the axial parts of the quark current  $\bar Q_i\gamma_\nu(1-\gamma^5)Q_j$ $ =V_\nu - A_\nu$, see \cite{Bondarenko:2018ptm}:
\begin{eqnarray}
  \langle h'_V(\epsilon,p')  |V_\mu| h(p)\rangle\! =\! {\rm i} g(q^2) \varepsilon_{\mu\alpha\sigma\rho}
  \epsilon^{*\alpha} (p\!+\!p')^{\sigma} (p\!-\!p')^{\rho}
  = {\rm i} 2 g(q^2) \varepsilon_{\mu\alpha\sigma\rho}
  \epsilon^{*\alpha} p^{'\,\sigma} p^{\rho} =   {\rm i} \mathbb{V}_\mu,\label{Dstar1}\\
  \langle h'_V(\epsilon,p')|  A_\mu | h(p)\rangle\! =\!
  f(q^2) \epsilon_\mu^*\! +
  a_+(q^2) (\epsilon^*\!\cdot\! p) (p\!+\!p')_\mu\! 
  +  a_-(q^2) (\epsilon^*\!\cdot\! p) (p\!-\!p')_\mu  = \mathbb{A}_\mu,\label{Dstar2}
\end{eqnarray}
and  $\epsilon_\mu$ is the polarization vector of the vector meson.

For $|M_{fi}|^2$ summarized over the helicities and  polarization states of the final particles, we have
\begin{equation}
\overline{|M_{fi}|^2} =4\theta_\alpha^2 G_F^2 |V_{ij}|^2 
\sum_{\lambda} R^{\mu\nu} 
[\mathbb{V}_\nu \mathbb{V}_\mu^* + \mathbb{A}_\nu \mathbb{A}_\mu^*+{\rm i}(\mathbb{A}_\nu \mathbb{V}_\mu^*-\mathbb{V}_\nu \mathbb{A}_\mu^*)],
\end{equation}
where $\lambda$ is the polarization state of the vector meson and
\begin{equation}
  R^{\mu\nu} = (q^\nu - k^\nu)k^\mu -(qk)g^{\nu\mu} + m_N^2g^{\nu \mu} + (q^\mu - k^\mu)k^\nu - {\rm i} q_ik_j\varepsilon^{ij\nu\mu}.
\end{equation}

Summation over the polarization states of the vector meson can be performed directly using the relation
\begin{equation}
    \sum_\lambda \varepsilon^*_\alpha (p') \varepsilon_\beta (p')=-\left(g_{\alpha\beta}-\frac{p'_\alpha p'_\beta}{m_{h^\prime_V}^2} \right).
    \end{equation}
We get
\begin{equation}
   \sum_\lambda R^{\mu\nu} \mathbb{V}_\nu \mathbb{V}_\mu^* 
   = 8g^2(q^2) 
   \bigl\{ m_B^2[(p'k)^2 - m_D^2 (qk) ] + m_D^2(pk)^2 + ( pp')[ -2 (p'k)(pk) + (pp')(qk)  ]  \bigr\},
\end{equation}
\begin{equation}
   \sum_\lambda R^{\mu\nu} {\rm i} (\mathbb{A}_\nu \mathbb{V}_\mu^*-\mathbb{V}_\nu \mathbb{A}_\mu^*)
    = 8 g(q^2) f(q^2)
    [  m_B^2 (p'k) + m_D^2 (pk) - (p'k + pk)(pp')   ],
\end{equation}
\begin{eqnarray}
    \sum_\lambda R^{\mu\nu}  \mathbb{A}_\mu \mathbb{A}_\nu^*=
   - R^2(m_N^2 - qk)\biggl(m_B^2 - \frac{(pp')^2}{m_D^2}\biggr) + \frac{2(kR)(kR - qR)}{m_D^2} +\nonumber\\+ 2f(q^2) 
   \bigl[(kQ)(qR) + (kR)(qQ-2kQ) + (RQ)(m_N^2 -qk)\bigr] +\nonumber\\+ f^2(q^2)
  \Biggl( \frac{2(p'k)(p'p-p'k)}{m_D^2} - m_N^2 + qk - 2p'k\Biggr),
\end{eqnarray}
 where
\begin{equation}
     R_\mu =a_+(q^2)(p+p')_\mu +  a_-(q^2)(p-p')_\mu,\quad
     Q_\nu=\frac{(pp')}{m_D^2} p'_\nu - p_\nu.
\end{equation}

Form-factors $f(q^2)$, $g(q^2)$, $a_\pm(q^2)$ can be found from the dimensionless linear combinations \cite{Ebert,Faustov,Melikhov}:
\begin{align}
 V^{h h'}(q^2) &= \left(m_h + m_{h'}\right) g^{h h'}(q^2),
 \\
 A_0^{h h'}(q^2) &= \frac{1}{2 m_{h'}}
 \Bigl(
 f^{h h'}(q^2) + q^2 a_-^{h h'}(q^2) + \left(m_h^2 - m_{h'}^2 \right) a_+^{h h'}(q^2)
 \Bigr),
 \\
 A_1^{h h'}(q^2) &= \frac{f^{h h'}(q^2)}{m_h + m_{h'}},
 \\
 A_2^{h h'}(q^2) &= - \left(m_h + m_{h'}\right) a_+^{h h'}(q^2),
\end{align}
that can be parameterized as
\begin{align}
 V^{h h'}(q^{2}) &= \frac{f^{h h'}_{V}}{1-q^{2}/(M^{h}_{V})^{2}[1-\sigma_{V}^{h h'} q^{2} / (M^{h}_{V})^{2} - \xi_{V}^{h h'} q^{4} / (M^{h}_{V})^{4}]},
 \label{eq:Vhh}
\\
 A_{0}^{h h'}(q^{2}) &= \frac{f^{h h'}_{A_{0}}}{1-q^{2}/(M^{h}_{P})^{2}[1-\sigma_{A_{0}}^{h h'} q^{2} / (M^{h}_{V})^{2} - \xi_{A_{0}}^{h h'} q^{4} / (M^{h}_{V})^{4}]},
 \label{eq:A0hh}
\\
 A_{1/2}^{h h'}(q^{2}) &= \frac{f^{h h'}_{A_{1/2}}}{1 - \sigma_{A_{1/2}}^{h h'} q^{2} / (M^{h}_{V})^2 - \xi_{A_{1/2}}^{h h'} q^{4} / (M^{h}_{V})^{4}}.
 \label{eq:A12hh}
\end{align}

The best fit values of parameters are given in papers~\cite{Ebert,Faustov,Melikhov}. The parameters $f$ and $\sigma$ are given in table  \ref{tab:vector-form factors-constants}. The parameters $\xi$ and the pole masses $M_V$, $M_P$ are given in table  \ref{tab:pole-masses}, where
$m_{D_{s}} = 1.969$, $m_{D_s^{*}} = 2.112$, $m_{B_{c}} = 6.275$, $m_{B_{c}^{*}}= 6.331$. 
The mass for $B_c^*$ was taken from theoretical prediction~\cite{Mathur}.

\begin{table}[t]
\noindent \caption{First part of the table with parameters of form-factors~(\ref{eq:Vhh}-\ref{eq:A12hh}) of $B$ and $D$ mesons decays into vector mesons~\cite{Ebert,Faustov,Melikhov}.}
    \label{tab:vector-form factors-constants}\vskip3mm\tabcolsep4.5pt
    \centering
    \begin{tabular}{|c|c|c|c|c|c|c|c|c|}
        \hline
        $h,h'$  &  $f_{V}^{hh'}$ &  $f_{A_{0}}^{hh'}$ &  $f_{A_{1}}^{hh'}$ &  $f_{A_{2}}^{hh'}$  &  $\sigma_{V}^{hh'}$  &  $\sigma_{A_{0}}^{hh'}$  &  $\sigma_{A_{1}}^{hh'}$  &  $\sigma_{A_{2}}^{hh'}$ 
        \\
        \hline
         $D, K^{*}$  & $1.03$ & $0.76$ & $0.66$ & $0.49$ & $0.27$ & $0.17$ & $0.30$ & $0.67$
        \\
        \hline
         $B, D^{*}$  & $0.76$ & $0.69$ & $0.66$ & $0.62$ & $0.57$ & $0.59$ & $0.78$ & $1.40$
      \\  \hline
    \end{tabular}
\end{table} 

\begin{table}[t]
\noindent\caption{Second part of the table with parameters of
          form-factors~(\ref{eq:Vhh}-\ref{eq:A12hh}) of $B$ and $D$ mesons decays into vector mesons~\cite{Ebert,Faustov,Melikhov}.}
    \label{tab:pole-masses}\vskip3mm\tabcolsep4.5pt
    \centering
    \begin{tabular}{|c|c|c|c|c|c|c|}
        \hline
        $h,h'$ & $\xi_{V}^{hh'}$ &  $\xi_{A_{0}}^{hh'}$ & $\xi_{A_{1}}^{hh'}$ & $\xi_{A_{2}}^{hh'}$ &  $M_{P}^{h}$ (GeV)  &  $M_{V}^{h}$ (GeV)  \\
        \hline
         $D, K^{*}$  & 0 & 0 & 0.20 & 0.16 & $m_{D_{s}}$ &$m_{D_s^{*}}$ \\
        \hline
         $B, D^{*}$  & 0 & 0 & 0 & 0.41 & $m_{B_{c}}$& $m_{B_{c}^{*}}$\\
        \hline
    \end{tabular}
\end{table}

\section{PDF for HNLs produced in 3-body decays}\label{sec:pdf_hnl}

In this paper, we consider  HNL production in 3-body decays of pseudoscalar  $B$ and $D$ mesons into  other 
pseudoscalar mesons, namely $B^\pm \rightarrow D^0 + \ell^\pm +N$ and $D^\pm \rightarrow K^0 +\ell^\pm + N$. 
Also  we consider  HNL production in 3-body decays of pseudoscalar  $B$ and $D$ mesons into   
vector mesons, namely $B^\pm \rightarrow D^*(2007)^0 + \ell^\pm + N$ and  $D^{\pm} \rightarrow K^*(892) + \ell^{\pm} + N$. 
Two cases of leptons $(\ell=e,\mu)$  in the final states are 
practically indistinguishable because of the small masses of electron and muon as compared with the masses of mesons in the reactions.

\begin{figure}[t]% figure* for wide figure, [h] [!] to change the placement
\centering
\includegraphics[width=0.9\textwidth]{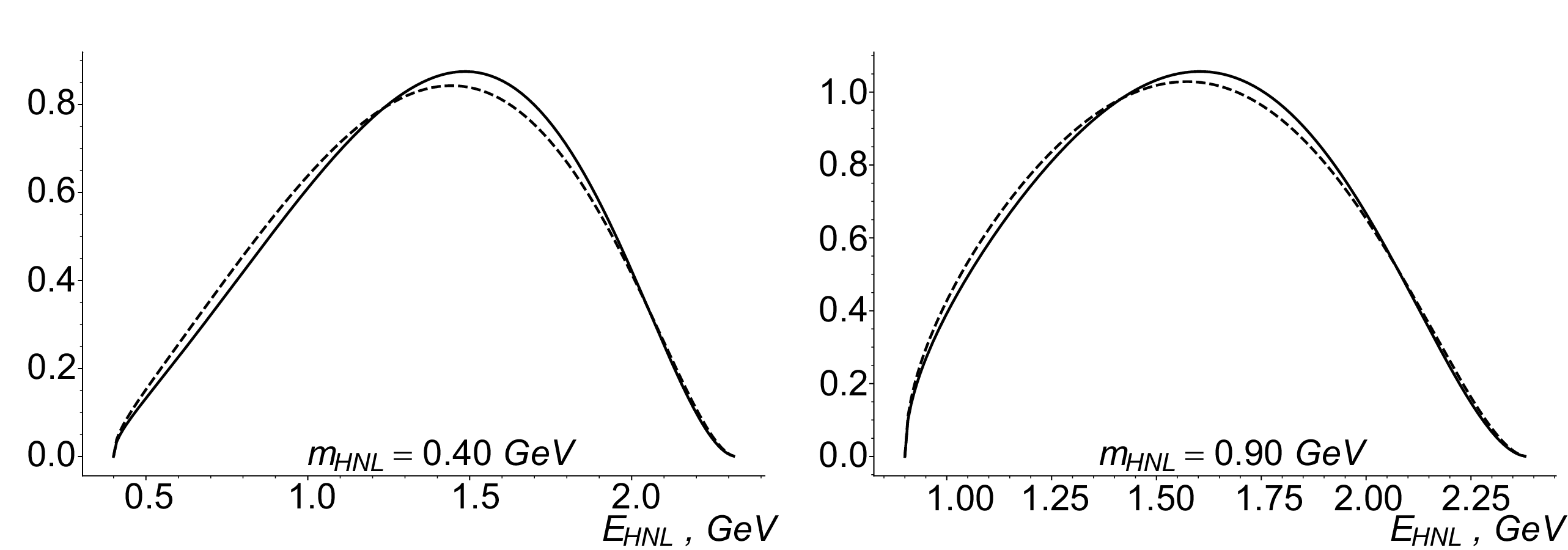}
\caption{Probability density functions for the energy of HNLs (in the rest frame of the initial meson) produced in the reaction $B^\pm \rightarrow D^0 + \ell^\pm + N$  computed exactly (solid line) and with the PYTHIA default matrix element for decays of $B$ meson \eqref{PythB} (dashed line).}\label{BDPythB}
\end{figure}

Let us compare the probability density function in the own frame of the initial meson for production of HNLs with
a certain energy ($E_N$) computed with the help of relation \eqref{pdf_def} for the exact matrix elements (see section \ref{decay:pseudoscalar} and section \ref{decay:vector}) and for the parton level PYTHIA default matrix elements, see \eqref{PythD}, \eqref{PythB}. 

\begin{figure}[b!]% figure* for wide figure, [h] [!] to change the placement
\centering
\includegraphics[width=0.9\textwidth]{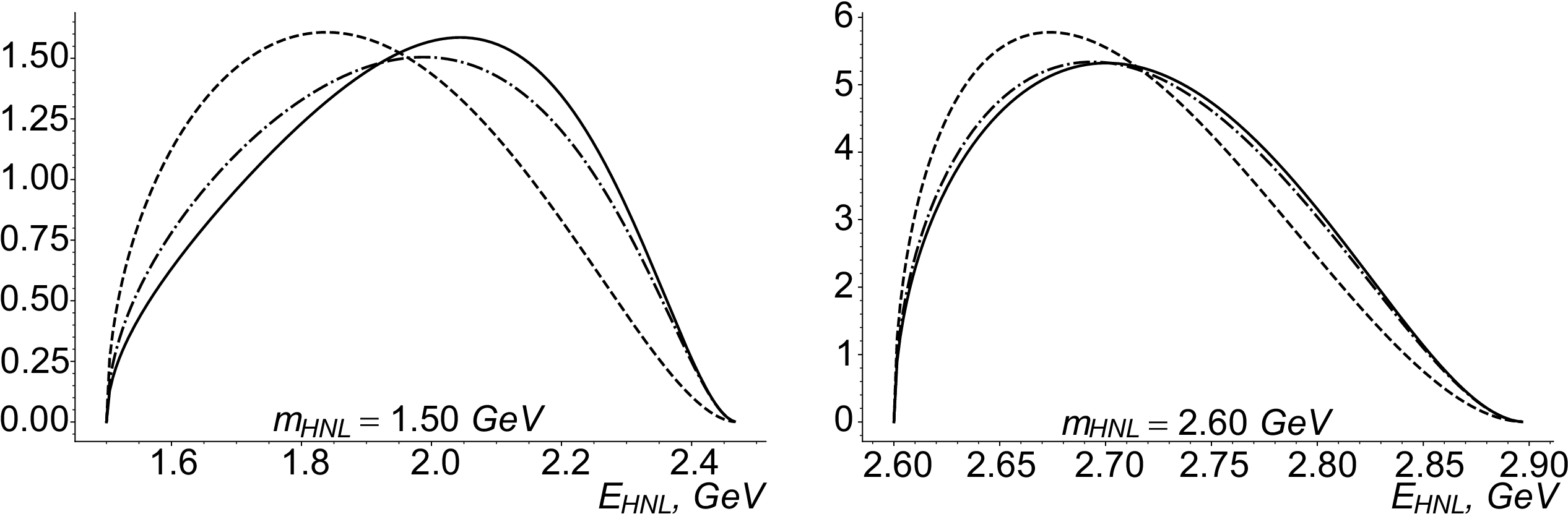}
\caption{Probability density functions for the energy  of HNLs (in the rest frame of the initial meson) produced in the reaction  $B^\pm \rightarrow D^*(2007)^0 +\ell^\pm + N$ computed exactly (solid line), with the PYTHIA default matrix element for decays of $B$ meson \eqref{PythB} (dashed line) and with the PYTHIA default matrix element for decays of $D$ meson \eqref{PythD} (dot-dashed line).}\label{BDstarPythB}
\end{figure}

We see that the PDF computed exactly for the reaction of HNL production in the decay of the pseudoscalar meson $B$  into the pseudoscalar meson $D^0$   is in good agreement with the PDF computed with the PYTHIA default matrix element for the decay of $B$ mesons \eqref{PythB}, see figure \ref{BDPythB}. 
But the PDF computed exactly for the reaction of the $B$ meson decay into a vector meson $B^\pm \rightarrow D^*(2007)^0 + \ell^\pm + N$  is in good agreement not with the PDF computed with the PYTHIA default matrix element for decays of $B$ mesons \eqref{PythB},
but with the PDF computed with the PYTHIA default matrix element for decays of $D$ mesons \eqref{PythD}, see figure \ref{BDstarPythB}.

\begin{figure}[t]% figure* for wide figure, [h] [!] to change the placement
\centering
\includegraphics[width=0.9\textwidth]{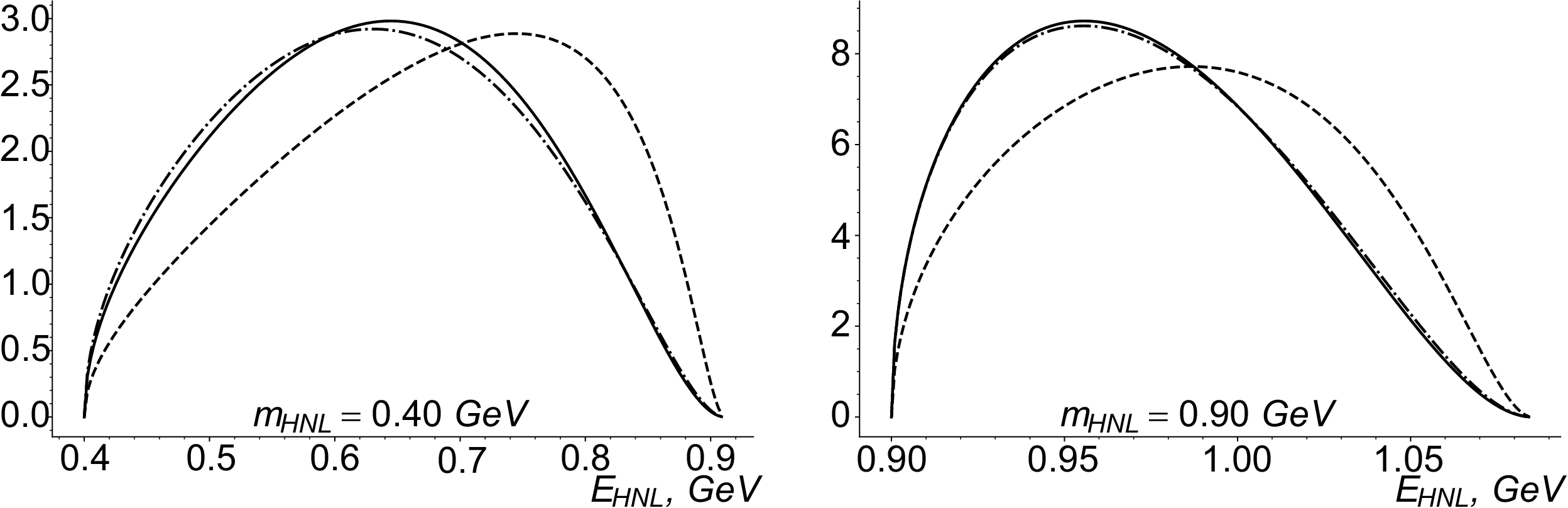}
\caption{Probability density functions for the energy  of HNLs (in the rest frame of the initial meson) produced in the reaction  $D^\pm \rightarrow K^0 +\ell^\pm + N$ computed exactly (solid line) and with the PYTHIA default matrix element for decays of $D$ meson \eqref{PythD} (dashed line) and with the PYTHIA default matrix element for decays of $B$ meson \eqref{PythB} (dot-dashed line).}\label{DKPythD}
\end{figure}

\begin{figure}[b!]% figure* for wide figure, [h] [!] to change the placement
\centering
\includegraphics[width=0.9\textwidth]{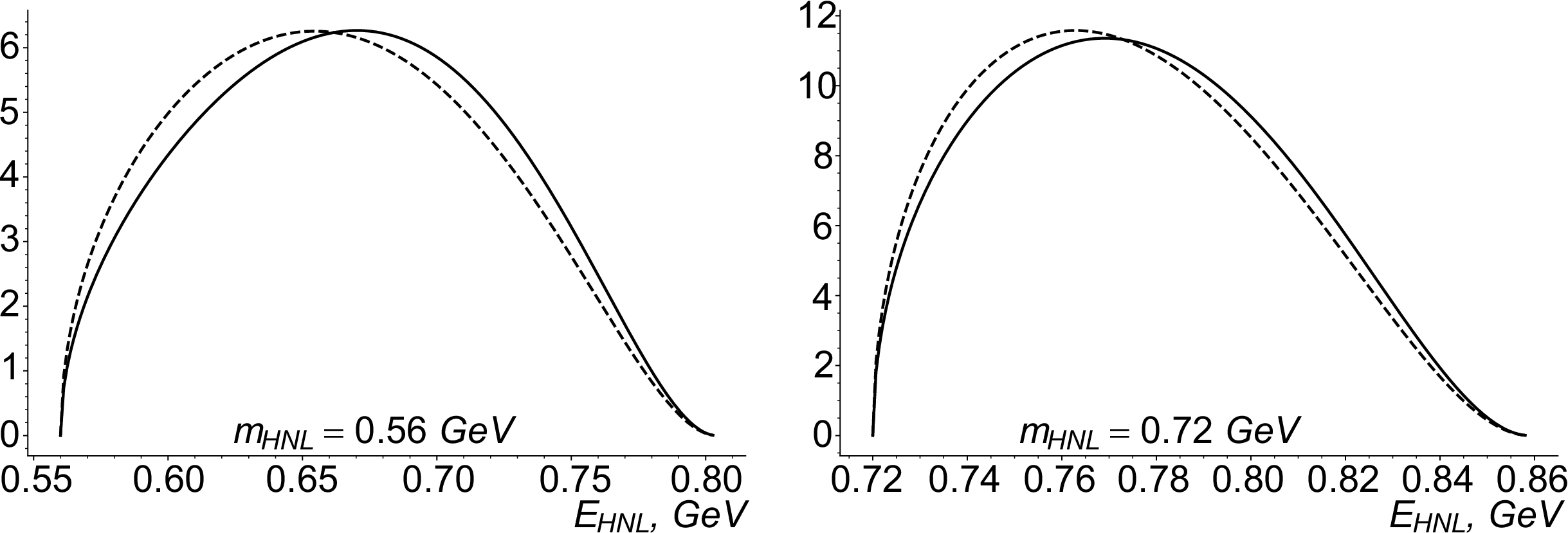}
\caption{Probability density functions for the energy  of HNLs (in the rest frame of the initial meson) produced in the reaction  $D^\pm \rightarrow K^*(892) +\ell^\pm + N$ computed exactly (solid line) and with the PYTHIA default matrix element for decays of $D$ meson \eqref{PythD} (dashed line).}\label{DKstarPythD}
\end{figure}

We see that the PDF computed exactly for the reaction of HNL production in the decay of the pseudoscalar meson $D$ into the pseudoscalar meson $K^0$ is in good agreement not with the PDF computed with the PYTHIA default matrix element of the $D$ meson decays  \eqref{PythD}, but with the PDF computed with the PYTHIA  default matrix element for decays of $B$ mesons \eqref{PythB}, see figure \ref{DKPythD}.
But the PDF computed exactly for the reaction of $D$ meson decay into a vector meson $D^\pm \rightarrow K^*(892) + \ell^\pm + N$  is in good agreement with the PDF computed with the PYTHIA default matrix element for decays of $D$ mesons \eqref{PythD}, see figure \ref{DKstarPythD}.

The result, at least for the initial $B$ and $D$ mesons,  can be summarized as follows. The parton level PYTHIA default matrix element \eqref{PythB} works well for a pseudoscalar meson 3-body decay into a pseudoscalar meson, a charged lepton and an HNL. The parton level PYTHIA default matrix element \eqref{PythD} works well for a pseudoscalar meson 3-body  decay into a vector meson, a charged lepton and an HNL.

An intriguing question is how the sufficiently large difference in the PDFs will affect the quantities necessary for calculating the sensitivity region, namely $\epsilon_{geom}$ and $P_{decay}$. We consider this question in section \ref{sec:egeom} and section \ref{sec:pdecay}.

\section{Distribution functions}\label{sec:dist_funcs} %\cite{shipPythia}

Data \textit{Datah} for distribution of
the produced $h$  mesons ($h$ stands for $B$ or $D$ mesons) in proton-target collisions (along axis $z$) in the SHiP experiment over the energy ($E_h$) and polar angle ($\theta_h$) values were kindly provided by the SHIP collaboration. These data were obtained by using a tuned PYTHIA 6.4, see \cite{shipPythia} for more details. In the following
calculations, a data array with $a_B=6.27\cdot 10^5$ elements of type
$(E_B,\theta_B)$  and a data array with $a_D=10^6$ elements of type
$(E_D,\theta_D)$ are used. 

\begin{figure}[t]% figure* for wide figure, [h] [!] to change the placement
\centering
\includegraphics[width=\textwidth]{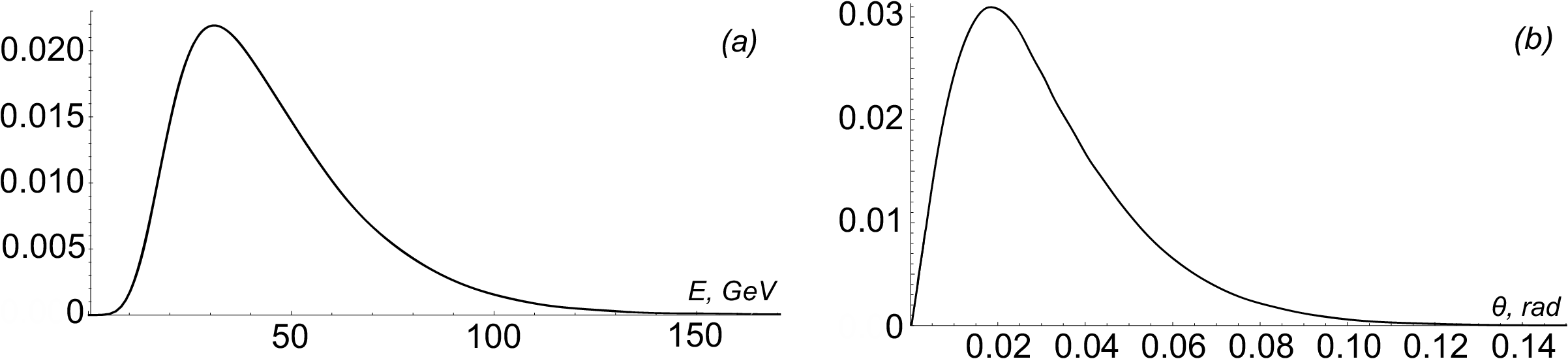}
\caption{Probability density functions for energy (\textit{a}) and polar angle (\textit{b}) in the laboratory frame for HNLs with mass $2.6$ GeV produced in the  reaction $B^\pm \rightarrow D^*(2007)^0 +\ell^\pm + N$.}\label{EN_Th_Dist_BDstar}
\end{figure}

We use these data  to calculate the distribution functions for the energy
($E_N^{lab}$) and the polar angle ($\theta_N^{lab}$) of the produced
HNLs in the laboratory frame with the help of Monte-Carlo simulation. Kinematic relations for HNLs between the laboratory and $h$ meson's own  reference frames are presented in appendix A.
Corresponding relations %\eqref{a26}, \eqref{a27}
contain dependence on
two parameters of the $h$ meson in the laboratory reference frame
(the energy $E_h$ and the polar angle $\theta_h$) and  four parameters of
the HNL in the own reference frame of the $h$ meson (the mass of the HNL $m_N$, the energy of the HNL $E_N^{cm}$, the polar and azimuth angles of the HNL
$\theta^{cm}_N$, $\varphi_N^{cm}$). 

It should be noted that in the
$h$ meson's own reference frame,  the energy 
of the HNL is defined by the energy probability distribution function \eqref{pdf_def}, but  directions of
motion of the produced HNL are equiprobable because of the
isotropic property of the $h$ meson decay.
So we take these  parameters
($y=\cos\theta^{cm}_N$ and $\varphi_N^{cm}$) with randomly chosen
values. A set of values ($E_h,\theta_h$) is  taken from the elements
of \textit{Datah} with a serial number that is taken in a random way.
Thus we receive  data (\textit{DataN})  in a form
$(E_N^{lab},\theta^{lab}_{N} )$.

Using the obtained data, we can now derive probability density functions
(PDF) and cumulative distribution functions (CDF) for energy and
polar angle for HNLs and for the initial $h$ mesons also. As an example, we present the PDFs for energy and polar angle (computed via the matrix element from section \ref{decay:vector}) in the 
laboratory reference frame for HNLs with mass 2.6 GeV produced in reaction
$B^\pm \rightarrow D^*(2007)^0 +\ell^\pm + N$, see figure
\ref{EN_Th_Dist_BDstar}.

Using the definition of the median  value as a value
corresponding to the cumulative distribution function equal to  1/2,
we get the median values of energy and polar angle for the
initial $B$  and $D$ mesons produced in the SHiP experiment, namely $\overline{E_B} \simeq 80 $ GeV, 
$\overline{\theta_B} \simeq 0.022$ and  $\overline{E_D} \simeq 16.5 $ GeV, 
$\overline{\theta_D} \simeq 0.022$.

For the produced HNLs the corresponding median values depend on their masses and the reaction of their production. 
As an example,  for an HNL with mass $2.6$ GeV produced in the reaction $B^- \rightarrow D^*(2007)^0 +e^- + N$, see   figure
\ref{EN_Th_Dist_BDstar},
the median values of energy and polar angle  in the laboratory reference frame are
$\overline{E_N} \simeq 40$ GeV and $\overline{\theta_N}
\simeq 0.026$. 

\section{Factor $\epsilon_{geom}$}\label{sec:egeom}

In this section we  consider the probability  of a produced
HNL to move toward the detector ($\epsilon_{geom}$). To do this,
we have to find the probability of the polar angle $\theta_N^{lab}$
of the produced HNL to be less than the angular size of the
detector e.g. for the SHiP experiment $\theta_{detector}=0.028$.  It is just the value of the cumulative
distribution function $(F_{N,\theta})$ for the polar angle of the
HNL  in the laboratory  frame
\begin{equation}\label{a29}
\epsilon_{geom}(m_N)=F_{N,\theta}(m_N)=\hspace{-0.8em}\int\limits_0^{\theta_{detector}}\hspace{-0.8em} pdf(m_N,\theta)d\theta,
\end{equation}
where $pdf(m_N,\theta)$ is the PDF on polar angle $\theta$ of the HNL.
Value of $\epsilon_{geom}$ depends on the
mass of the HNL and the matrix element for its production.

\begin{figure}[t]% figure* for wide figure, [h] [!] to change the placement
\centering
\includegraphics[width=\textwidth]{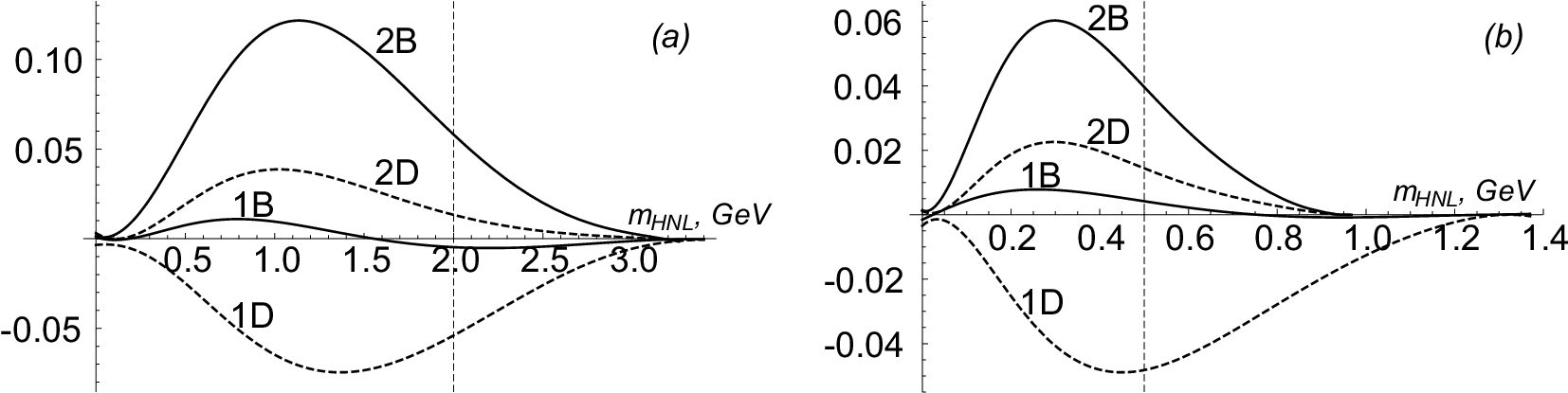}
\caption{We present ratio $\frac{\epsilon_{geom}^{PYTH}-\epsilon_{geom}}{\epsilon_{geom}}$ for the produced HNLs. 
Figure (\textit{a}) corresponds to the $B$ meson decays (lines 1B and 1D:   $ B^\pm \rightarrow D^0 + \ell^\pm + N $, lines 2B and 2D:  $ B^\pm \rightarrow D^*(2007)^0 + \ell^\pm + N $).
Figure (\textit{b}) corresponds to the $D$ meson decays (lines 1D and 1B: $ D^\pm \rightarrow K^0 + \ell^\pm + N $, lines 2D and 2B: $D^\pm \rightarrow K^*(892) + \ell^\pm + N$). 
Dashed line corresponds to computations via the PYTHIA matrix element \eqref{PythD}. Solid line corresponds to computations via the PYTHIA matrix element \eqref{PythB}. We are only interested in the area to the right of the vertical dashed line. }\label{pgeom_ratio}
\end{figure}

The result of our computations can be presented in the following way. 
Factor $\epsilon_{geom}$ computed exactly (with the help of matrix elements presented in section \ref{decay:pseudoscalar} and section \ref{decay:vector}) for the reactions $B^\pm \rightarrow D^0 + \ell^\pm +N$ and $D^\pm \rightarrow K^*(892) +\ell^\pm + N$ is in good agreement
(the relative difference is of order  $1\%-2\%$ in the area to the right of the vertical dashed line) 
with factor $\epsilon_{geom}$ computed with the help of the parton level PYTHIA default matrix elements for decays of $B$  and $D$ mesons correspondingly, see line 1B of figure \ref{pgeom_ratio}(a) and line 2D of figure \ref{pgeom_ratio}(b). But for the reactions $B^\pm \rightarrow D^*(2007)^0 + \ell^\pm + N$ and $D^\pm \rightarrow K^0 + \ell^\pm + N$ 
it is more 
preferable to use the parton level PYTHIA default matrix elements for decays of $D$  and $B$ mesons correspondingly,
see line 2D of figure \ref{pgeom_ratio}(a) and line 1B of figure \ref{pgeom_ratio}(b).  
It should be noted that a similar situation was  for computations of PDFs in section \ref{sec:pdf_hnl}. 
 We emphasize that in figure \ref{pgeom_ratio}(b) the experimentally interesting area is $m_{HNL}\gtrsim 0.5$ GeV (to the right of the vertical dashed line) and in figure \ref{pgeom_ratio}(a) it is $m_{HNL}\gtrsim 2$ GeV (to the right of the vertical dashed line),
see section \ref{intro}.

As one can see from figure \ref{pgeom_ratio} the probabilities of a produced HNL to move towards the detector coincide ($\epsilon_{geom}^{PYTH}=\epsilon_{geom}$)
for an HNL with the maximum kinetically allowed value of the mass. 
This is how it should be.
In the case when an HNL has the maximum value of mass, the initial hadron in its own reference frame decays into three motionless particles. All these particles in the laboratory frame will move in the  direction of the initial meson regardless of the matrix element type.

\section{Factor $P_{decay}$}\label{sec:pdecay}

The number of the produced HNLs that can be detected during
the  period of the SHIP operation is defined by \eqref{Ns0}. This
relation allows us to find the region of parameters  $(m_N,U_\alpha)$,
with which  HNLs can be detected. 

Behavior of the lower bound  of this region for coupling constant $(\theta)$
can be estimated analytically. In this case arguments of the
exponents in $P_{decay}$ \eqref{a31} are small and approximation
$e^x\simeq1+x$ can be used and we get
\begin{equation}\label{PdecaySmall}
P_{decay}\approx \Delta L\cdot \Gamma \cdot X(E_N/m_N),
\end{equation}
where $X(u)=(u^2 - 1)^{-1/2}$. 

\begin{figure}[t!]% figure* for wide figure, [h] [!] to change the placement
\centering
\includegraphics[width=\textwidth]{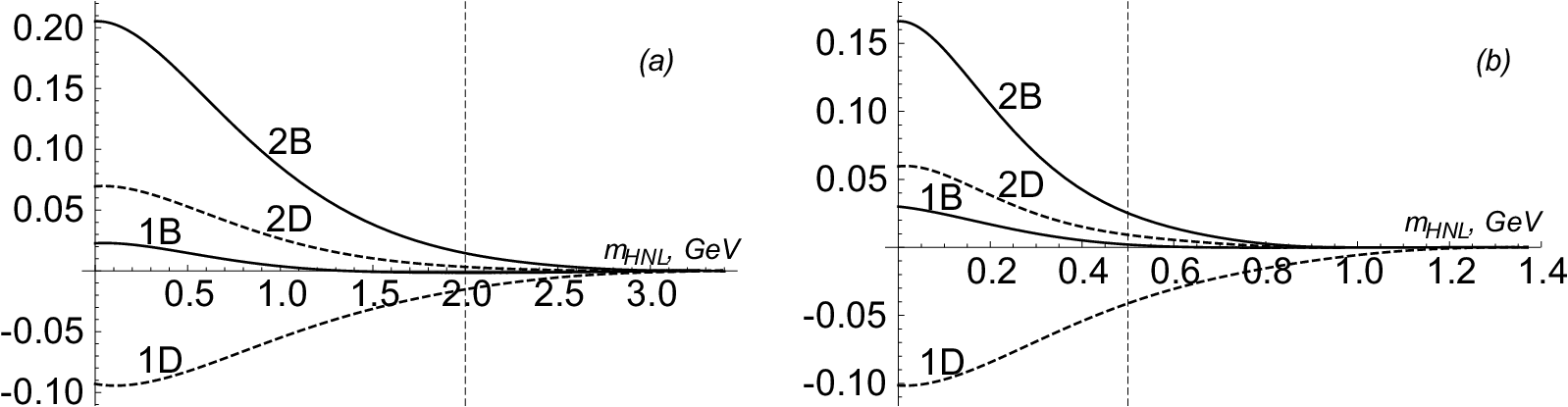}
\caption{We present ratio $\frac{P_{decay}^{PYTH}-P_{decay}}{P_{decay}}$for the produced HNLs. 
Figure (\textit{a}) corresponds to the $B$ meson decays (lines 1B and 1D:   $ B^\pm \rightarrow D^0 + \ell^\pm + N $, lines 2B and 2D:  $ B^\pm \rightarrow D^*(2007)^0 + \ell^\pm + N $).
Figure (\textit{b}) corresponds to the $D$ meson decays (lines 1D and 1B: $ D^\pm \rightarrow K^0 + \ell^\pm + N $, lines 2D and 2B: $D^\pm \rightarrow K^*(892) + \ell^\pm + N$). 
Dashed line corresponds to computations via the PYTHIA matrix element \eqref{PythD}. Solid line corresponds to computations via the PYTHIA matrix element \eqref{PythB}. We are only interested in the area to the right of the vertical dashed line.}\label{pdecay_ratio}
\end{figure}

With the help of \eqref{PdecaySmall} we can easily obtain the ratio of $P_{decay}$ computed for the exact matrix elements (see section \ref{decay:pseudoscalar} and section \ref{decay:vector}) and for the matrix elements in the PYTHIA approximation, see \eqref{PythD} and \eqref{PythB}:
\begin{equation}\label{PdecayRatio}
\frac{P_{decay}^{PYTH}}{P_{decay}}=\frac{{\tilde X}^{PYTH}(E_N/m_N)}{{\tilde X}(E_N/m_N)},
\end{equation}
where ${\tilde X}$ is the median value of the corresponding function $X$. The median value of the function $X$ was found by computing the median of the set of
the $X$-function values produced by Monte Carlo simulations, see section \ref{sec:dist_funcs}. 

The results of our computations are similar to the results of the 
analysis of the $\epsilon_{geom}$ factor. 
Factor $P_{decay}$ computed exactly (with the help of the matrix elements presented in section \ref{decay:pseudoscalar}) for the reactions $B^\pm \rightarrow D^0 + \ell^\pm +N$ is in good agreement (the relative difference is about  1$\%$  for $m_N\gtrsim2$ GeV) with factor $P_{decay}$ computed with the help of the parton level PYTHIA default matrix elements \eqref{PythB} for decays of $B$ mesons, see line 1B of figure \ref{pdecay_ratio}(a). Factor $P_{decay}$ computed exactly (with the help of the matrix elements presented in section \ref{decay:vector}) for the reactions $D^\pm \rightarrow K^*(892) +\ell^\pm + N$ is in  agreement (the relative difference less than 3$\%$ for $m_N\gtrsim0.5$ GeV) with factor $P_{decay}$ computed with the help of the parton level PYTHIA default matrix elements \eqref{PythD} for decays of $D$ mesons, see line 2D of figure \ref{pdecay_ratio}(b). 

It should be noted that for the reactions $B^\pm \rightarrow D^*(2007)^0 + \ell^\pm + N$ and $D^\pm \rightarrow K^0 + \ell^\pm + N$
it is more 
preferable to use the parton level PYTHIA default matrix elements for decays of $D$  and $B$ mesons correspondingly, see line 2D of figure \ref{pdecay_ratio}(a) and line 1B of figure \ref{pdecay_ratio}(b).  
As in the case with $\epsilon_{geom}$ we are only interested in the area  to the right of the vertical dashed line.

\section{Conclusions}\label{conclusions}

\vspace{-1em}

There are some indisputable phenomena that point to the fact that the SM has to be modified and complemented by a new particle (particles). We are sure that new physics exists, but we do not know where to search for it. There are many theoretical possibilities  to modify the SM, namely by scalar, pseudoscalar, vector, pseudovector, or fermion particles of new physics. These particles may be substantially heavier than the energy scale of the present colliders. However, they may also be light (with mass less than the electroweak scale) and feebly interact with the SM particles. 

In this paper, we consider the HNL extension of the SM. We analyzed the relevance  of using the parton level PYTHIA
default matrix elements without additional tuning for  describing GeV-scale HNL production 
in the most important 3-body decays of $B$, $D$ mesons  that is a topical question for construction of the sensitivity region for the experimental search for HNLs. 
Our study was driven by the use of such an approximation in the SHiP collaboration paper \cite{Ahdida}.
We consider this question concerning the SHiP experiment, but our results are also applicable to other intensity frontier experiments.

Some general conclusions can now be drawn. The computations of the 3-body decays of $\tau$ leptons with HNL production in  the PYTHIA approximation coincide with the exact computations, but  the parton level PYTHIA default matrix elements for describing the 3-body decays of mesons are just similar to the  matrix elements for free quark electroweak decays.
Despite this, we have shown that this PYTHIA approximation has the right to be used for construction of the sensitivity region for the experimental search for HNLs, provided one uses the suitable parton level PYTHIA default matrix elements.

We consider the case of 3-body decays of $B$ and $D$ mesons into a light meson, an HNL and either an electron or a muon.
These two cases of leptons in the final state are 
practically indistinguishable because of the small electron and muon masses as compared with the masses of mesons in the reactions. Reactions with $\tau$ leptons in the final state are either forbidden (for the decays of $D$ mesons) or ineffective (as it was pointed in the Introduction HNL production from $B$ meson decays can be neglected for HNLs with masses $m_N \lesssim 2$ GeV).

As it was shown in section \ref{sec:strategy},  explicit form of matrix elements for HNL production in a
3-body decay affects factors $\epsilon_{geom}$ and $P_{decay}$ (probabilities of the produced HNL to move toward the detector and to decay inside the vacuum tank before the detectors correspondingly)  
in the relation  \eqref{Ns0} that defines the sensitivity region of the intensity frontier experiments. 

In section \ref{sec:egeom} and section \ref{sec:pdecay} we conducted a detailed analysis of factors $\epsilon_{geom}$ and $P_{decay}$. Summing up the results of this analysis, one can conclude that
for the description of HNL production in a 3-body decay of a pseudoscalar meson into another pseudoscalar meson ($B^\pm \rightarrow D^0 +\ell^\pm + N$ and $D^\pm \rightarrow K^0 +\ell^\pm + N$) 
the parton level PYTHIA matrix element \eqref{PythB}  is better to use.   For the description of HNL production in a 3-body decay of a pseudoscalar meson into a vector meson ($B^\pm \rightarrow D^*(2007)^0 +\ell^\pm + N$ and $D^\pm \rightarrow K^*(892) +\ell^\pm + N$) 
the parton level PYTHIA matrix element \eqref{PythD} is better to use.

\begin{figure}[t!]% figure* for wide figure, [h] [!] to change the placement
\centering
\includegraphics[width=\textwidth]{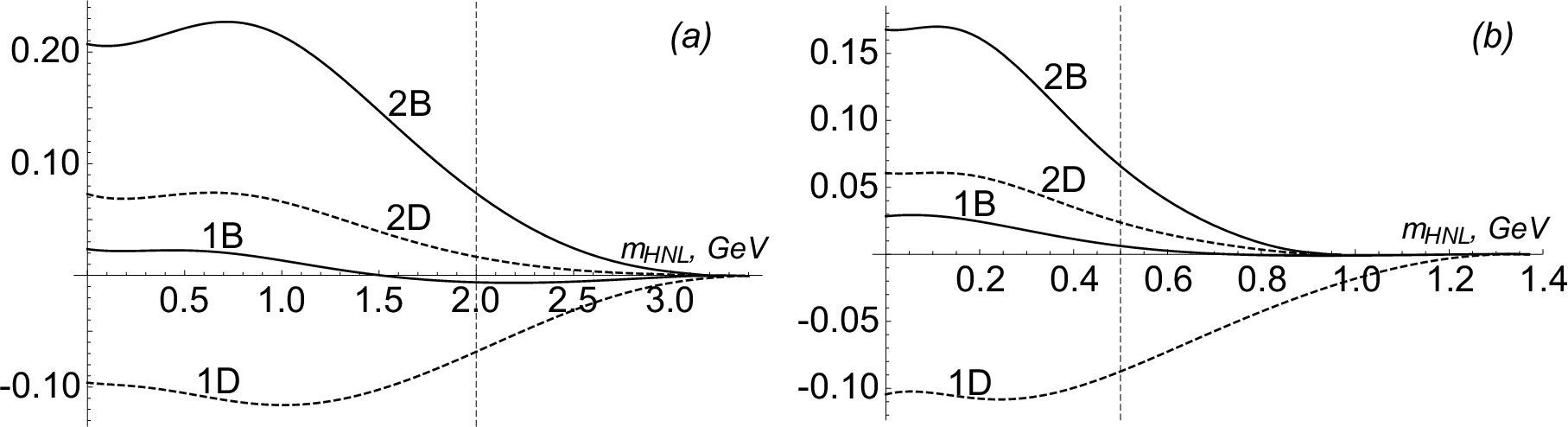}
\caption{\small We present ratio $\frac{\epsilon_{geom}^{PYTH}P_{decay}^{PYTH}-\epsilon_{geom}P_{decay}}{\epsilon_{geom}P_{decay}}$ for HNLs produced in the 3-body decays of mesons for the default PYTHIA matrix elements. Figure (\textit{a}) corresponds to the $B$ meson decays (lines 1B and 1D:   $ B^\pm \rightarrow D^0 + \ell^\pm + N $, lines 2B and 2D:  $ B^\pm \rightarrow D^*(2007)^0 + \ell^\pm + N $).
Figure (\textit{b}) corresponds to the $D$ meson decays (lines 1D and 1B: $ D^\pm \rightarrow K^0 + \ell^\pm + N $, lines 2D and 2B: $D^\pm \rightarrow K^*(892) + \ell^\pm + N$). 
Dashed line corresponds to computations via the PYTHIA matrix element \eqref{PythD}. Solid line corresponds to computations via  the PYTHIA matrix element \eqref{PythB}. We are only interested in the area to the right of the vertical dashed line.\vspace{-0.5em}}\label{error}
\end{figure}

Actually, it is more important to analyze the product $\epsilon_{geom}\cdot P_{decay}$, which is included in the relation \eqref{Ns0} that defines the sensitivity region for HNLs. Considering other factors in \eqref {Ns0} fixed, this allows us to obtain the total error of the quantity $N_ {det}$, which directly defines the accuracy of the sensitivity region construction. We demonstrate the discrepancies for $\epsilon_{geom}\cdot P_{decay}$ in figure \ref{error}.
 We emphasize that in figure \ref{error}(b) the experimentally interesting area is $m_{HNL}\gtrsim 0.5$ GeV (region to the left of the vertical dashed line is almost completely closed by experiments \cite{Alekhin, Bondar}) and in figure \ref{error}(a) the interesting area is $m_{HNL}\gtrsim 2$ GeV, where the production of HNLs in $B$ mesons decays is effective, see section \ref{intro}.

One can see that if we use default matrix element \eqref{PythB} for both the considered decays of $B$ meson (into $D^*(2007)$ as well as into $D^0$), we get quite a large maximum discrepancy $\simeq 7.3\%$ for the reaction
$B^\pm \rightarrow D^*(2007)^0 +\ell^\pm + N$, see line 2B of figure \ref{error}(a). 
Similarly, if we use default matrix element \eqref{PythD} for decays of $D$ meson into $K^0$ as well as into $K^*(892)$
we get quite a large maximum discrepancy $\simeq 8.8\%$ for reaction $D^\pm \rightarrow K^0 +\ell^\pm + N$, see
line 1D of figure \ref{error}(b). This is just the case implemented by the SHiP collaboration.

We come to the conclusion that if, for some reasons, under computations for HNL production in 3-body decays of mesons only default matrix elements are used, without additional tuning, it must be done in a more reasonable way.
The most suitable choice of the parton level PYTHIA default matrix elements \eqref{PythD}, \eqref{PythB}  is as shown by lines 1B and 2D for  figure
\ref{error}(a) and figure \ref{error}(b). With this choice, one can get, among all the considered $B$ and $D$ meson decays, the smallest difference with the exact matrix element for the reaction $D^\pm \rightarrow K^0 +\ell^\pm + N$ (the relative difference less than $0.6\%$), while the largest irremovable difference  is for the reaction $D^\pm \rightarrow K^*(892) +\ell^\pm + N$ (the relative difference less than $2.4\%$).

To summarize, it can be noted at least for the initial $B$ and $D$ mesons that the parton level PYTHIA default matrix element \eqref{PythB} works well for a pseudoscalar meson 3-body decay into a pseudoscalar meson, a charged lepton and an HNL. The parton level PYTHIA default matrix element \eqref{PythD} works well for a pseudoscalar meson 3-body  decay into a vector meson, a charged lepton and an HNL.  The interesting question of the possibility of  generalizing this statement for  other  initial  meson  states requires further investigation. 

It is important to note that the results of our research apply not only to physics beyond the Standard Model but also for the SM processes of semileptonic decays of mesons, 
see figure \ref{error} at zero HNL mass.

In no case should our research be considered as a call to use only the parton level
PYTHIA default matrix elements instead of accurate calculations using additional tuning for PYTHIA. We just checked the possibility and correctness of using  the default matrix elements approach.

\appendix
\setcounter{equation}{0}
\renewcommand{\theequation}{A.\arabic{equation}}
\section{Useful kinematic relations  for the
HNLs in the different reference frames}\label{appendix}

Let us consider an $h$ meson with mass $m_h$ and a given 4-momentum
$(E_h,\,{\vec p}_h)$  in the laboratory reference frame. Its velocity in this
reference frame (the origin of coordinates is at the point of proton-target
collisions, axis $z$ is directed to the center of detector) is
\begin{equation}\label{a8}
{\vec V}_h^{\,lab}=|\vec V_h^{\,lab}|\,{\vec e}_h^{\,\,lab},
\end{equation}
where
\begin{equation}\label{a8a}
|\vec V_h^{\,lab}|=|\vec p_h|\left(m_h^2+\vec
p{\,}_h^2\right)^{-1/2},\quad 
{\vec
e}_h^{\,\,lab}=(\sin\theta_h\cos\varphi_h,\sin\theta_h\sin\varphi_h,\cos\theta_h).
\end{equation}
$\theta_h$ and $\varphi_h$ are the  polar  and azimuth angles of the
$h$ meson's velocity vector in spherical coordinate system.

The absolute value of the
HNL's velocity in the center-of-mass system of the produced particles (own reference frame of the initial
$h$ meson) is
\begin{equation}\label{a12}
\vec V^{\,cm}_N=|\vec V^{cm}_N|\,{\vec e}^{\,\,cm}_N,
\end{equation}
where
\begin{equation}\label{a13}
|\vec V^{\,cm}_N|=|\vec p_{N,cm}|\left(m_N^2+\vec
p{\,}^2_{N,cm}\right)^{-1/2},\quad
{\vec
e}^{\,\,cm}_N\!=\!(\sin\theta_N^{cm}\cos\varphi_N^{cm},\sin\theta_N^{cm}\sin\varphi_N^{cm},\cos\theta_N^{cm}).
\end{equation}
$\theta_N^{cm}$ and $\varphi_N^{cm}$ are the  polar  and azimuth
angles of the HNL's velocity vector  in the center-of-mass
 system. It should be noted that in the center-of-mass
 system of the produced particles the decay of the $h$ meson is isotropic and the angles
$\theta_N^{cm}$ and $\varphi_N^{cm}$ can be arbitrary.

To find the value and the direction of the HNL's velocity in the
laboratory reference frame we have to find the components of $\vec V^{cm}_N$ that
are parallel and perpendicular to the direction of ${\vec
V}_h^{lab}$, namely $\vec V^{cm}_{N,||}$ and $\vec V^{cm}_{N,\bot}$.

It's obvious that $\vec V^{\,cm}_{N,||}=  V^{\,cm}_{N,||}\,{\vec
e}_h^{\,\,lab}$, where
\begin{equation}\label{a14}
 V^{\,cm}_{N,||}=|\vec V^{\,cm}_{N}|\,{\vec e}_h^{\,\,lab}\cdot
\vec e^{\,\,cm}_{N}=|\vec
V^{\,cm}_N|(\sin\theta_N^{\,cm}\sin\theta_h\cos(\varphi_h-\varphi_N^{\,cm})+
\cos\theta_N^{\,cm}\cos\theta_h)
\end{equation}
is the projection of vector $\vec V^{\,cm}_{N}$ on the direction ${\vec
e}_h^{\,\,lab}$.  Its value can be either positive or negative.

Consider now vector $\vec V^{cm }_{N,\bot}= |\vec V^{cm}_{N,\bot}|
\,\vec e^{\,\,cm}_{N,\bot}$, where
\begin{equation}\label{a14a}
|\vec  V^{cm}_{N,\bot}|=|\vec V^{\,cm }_{N}|\,|\vec
e^{\,\,cm}_{N}\times{\vec e}_h^{\,\,lab}|.
\end{equation}
Vector $\vec V^{cm}_{N,\bot}$ has to lie in the plane of vectors
$\vec V^{cm}_N$ and ${\vec V}_h^{lab}$. The equation of this plane is $
\vec n\cdot(\vec r-\vec r_0)=0 $, where $\vec n$ is the normal vector of
the plane $\vec n= {\vec e}^{\,\,cm}_N \times {\vec e}_h^{\,\,lab}$
and $\vec r-\vec r_0$  is vector lying in the  plane. Radii $\vec r $ and
$\vec r_0$ are meant to be taken with a base in the center of
target. We put ${\vec r}_0$ (point of the $h$ meson decay)   to be a
zero vector, because the distance between the point of production of meson and the point of its
decay is very small compared with the distance from the target to the detector.
As the vector $\vec r$ we can use the
 unit vector of direction  $\vec
e^{\,\,cm}_{N,\bot}$.

Components of the unit vector  $\vec
e^{\,\,cm}_{N,\bot}=(\tilde\alpha,\tilde\beta,\tilde\gamma)$ have to
satisfy the following equations%\vspace{-1.5em}
\begin{equation}\label{a18}
\left\{
\begin{array}{l}
 {\vec e}_h^{\,\,lab}\cdot \vec e^{\,\,cm}_{N,\bot}=\tilde\alpha \sin\theta_h\cos\varphi_h+\tilde\beta \sin\theta_h\sin\varphi_h+\tilde\gamma \cos\theta_h=0,\\
 n_x \tilde\alpha+n_y \tilde\beta+n_z \tilde\gamma=0,\\
 \tilde\alpha^2+\tilde\beta^2+\tilde\gamma^2=1.
\end{array}\right.
\end{equation}%\vspace{-0.5ex}
One can get the solution  in the form $\vec e^{\,\,cm}_{N,\bot}=\vec
N/{|\vec N|}$, where $\vec N=(\alpha,\beta,\gamma)$,
\begin{equation}\label{a19}\begin{aligned}
&\alpha=\cos\theta_N^{cm} \cos\theta_h\sin\theta_h \cos\varphi_h -\\&\qquad\qquad-
\sin\theta_N^{cm} \bigl(\cos^2\theta_h
\cos\varphi_N^{cm} + \sin^2\theta_h \sin(\varphi_h - \varphi_S^{cm})
\sin\varphi_h
 \bigr),\\&
\beta=\cos\theta_N^{cm} \cos\theta_h\sin\theta_h \sin\varphi_h  -\\&\qquad\qquad-
\sin\theta_N^{cm} \bigl(\cos^2\theta_h
\sin\varphi_N^{cm} - \sin^2\theta_h \sin(\varphi_h - \varphi_N^{cm})
\cos\varphi_h
 \bigr),\\&
\gamma=\sin\theta_h \bigl(\cos\theta_h \sin\theta_N^{cm
}\cos(\varphi_h - \varphi_N^{cm})-\sin\theta_B\cos\theta_N^{cm}
\bigr)
\end{aligned}\end{equation}
and the normalized coefficient $|\vec N|$ is equal to  $|\vec
e^{\,\,cm}_{N}\times{\vec e}_h^{\,\,lab}|$. The obtained solution has
ambiguity, because the system of equations \eqref{a18} is invariant
under simultaneous change of sign of all the vector's components. This
ambiguity can be removed under the following condition. If scalar product
$\vec N\cdot \vec V^{cm}_N$ is positive the sign of vector $\vec N$
is correct, otherwise the sign of vector $\vec N$ has to be changed.

So, vector $\vec V^{cm}_{N,\bot}$ is simply defined as
\begin{equation}\label{a20}
\vec V^{cm}_{N,\bot}=\alpha |\vec V^{cm}_{N,\bot}|\,\vec e^{\,\,cm}_{N,\bot}
=\alpha |\vec V^{\,cm}_{N,\bot}|\,\vec N/|\vec N|,
\qquad\alpha={\rm sgn}[\vec N\cdot {\vec e}{\,\,}^{cm }_N].
\end{equation}

Now we can find the value and the direction of the HNL's velocity in the
laboratory reference frame
\begin{align} &
 V_{N,||}^{lab}=\frac{|\vec V_h^{lab}|+ V_{N,||}^{cm} }{1+
V_{N,||}^{cm}\,|\vec V_h^{lab}|},
\quad{\vec V}_{N,||}^{lab}= V_{N,||}^{lab}\,{\vec e}_h^{\,\,lab}\label{a21}\\
& |\vec V^{lab}_{\bot}|=|\vec V^{cm}_{N,\bot}|\frac{\sqrt{1-|\vec
V_h^{lab}|^2} }{1+\vec V_{N,||}^{cm} |\vec V_h^{lab}|},
\quad\vec V^{lab}_{N,\bot} =\alpha |\vec V^{cm}_{N}|\vec N,\label{a22}
\\
& |\vec V_{N}^{lab}|=\sqrt{|\vec V_{N,||}^{lab}|^2 +|\vec
V_{N,\bot}^{lab}|^2}. \label{a23}
\end{align}
The components of the HNL's  velocity vector $\vec V_{N}^{lab}$ are
\begin{equation}\label{a26}\begin{aligned}
&(\vec V_{N}^{lab})_x= V_{N,||}^{lab}({\vec
e}_h^{\,\,lab})_x+\alpha |\vec V^{\,cm
}_{N}|(\vec N)_x=|\vec V_{N}^{lab}|\sin{\theta}^{lab}_N\cos\varphi^{lab}_N ,\\&
 (\vec V_{N}^{lab})_y= V_{N,||}^{lab}({\vec e}_h^{\,\,lab})_y+\alpha |\vec V^{\,cm}_{N}|(\vec N)_y=
|\vec V_{N}^{lab}|\sin\theta^{lab}_N\sin\varphi^{lab}_N,\\&
(\vec V_{N}^{lab})_z= V_{N,||}^{lab}({\vec
e}_h^{\,\,lab})_z+\alpha |\vec V^{\,cm }_{N}|(\vec N)_z=|\vec
V_{N}^{lab}|\cos\theta^{lab}_N,
\end{aligned}\end{equation}
where $({\vec e}_h^{\,\,lab})_i$ and $(\vec N)_i$ are the components
of vectors \eqref{a8a}, \eqref{a19} and $\theta^{lab}_N$,
$\varphi^{lab}_N$ are the angles of the HNL's velocity vector in
the laboratory reference frame. The energy of the HNL in the laboratory reference frame is
\begin{equation}\label{a27}
E_{N}^{lab}=m_N \left( 1-|\vec V_{N}^{lab}|^2 \right)^{-1/2}.
\end{equation}

In this paper we  assume that the experiment facility  has a cylindrical
symmetry. Therefore, the azimuth angle of the $h$ meson in the laboratory
reference frame can be set to zero ($\varphi_h=0$). The energy
$E_N^{lab}$ and the direction of the HNL's  velocity
$(\theta_N^{lab},\varphi_N^{lab})$ in the laboratory reference
frame are defined only by six parameters: the energy $E_h$ and the polar
angle $\theta_h$ of the $h$ meson in the laboratory reference frame,
 the two angles ($\theta_N^{cm}$ and $\varphi_N^{cm}$), the mass $m_N$ and the energy $E_N^{cm}$ of the produced HNL in the center-of-mass  system of the produced
particles (own reference frame of the $h$ meson).

\section*{Acknowledgments}
The authors are grateful to Alexey Boyarsky for statement of the problem and to Kyrylo Bondarenko for useful discussion and helpful comments.

\end{document}